\documentclass[seceq]{ptptex}

\usepackage{graphicx}

\newcommand{\be}{\begin{equation}}
\newcommand{\ee}{\end{equation}}
\newcommand{\bea}{\begin{eqnarray}}
\newcommand{\eea}{\end{eqnarray}}

\date{\today}

\title{
Cosmic structures via Bose Einstein condensation and its collapse
}

\author{Takeshi \textsc{Fukuyama}$^{\dag }$,
Masahiro \textsc{Morikawa}$^{ * } $ and Takayuki \textsc{Tatekawa}
$^{\circ}$
}

\inst{
$^{\dag }$Department of Physics, Ritsumeikan University, Kusatsu, Shiga,
525-8577 Japan \newline
Department of Physics, Yonsei University, Seoul 120-749, Korea

$^{ * } $Department of Physics, Ochanomizu University, 2-1-1 Otsuka,
 Bunkyo, Tokyo, 112-8610 Japan

$^{\circ }$The center for Continuing Professional Development, \ Kogakuin
University, 1-24-2 Nishi-shinjuku, Shinjuku, Tokyo 163-8677 Japan\newline
Advanced Research Institute for Science and Engineering, Waseda University,
3-4-1 Okubo, Shinjuku, Tokyo 169-8555 Japan
}

\abst{
We develop our novel model of cosmology based on the Bose-Einstein
condensation. This model unifies the Dark Energy and the Dark Matter, and
predicts multiple collapse of condensation, followed by the final
acceleration regime of cosmic expansion. We first explore the generality of
this model, especially the constraints on the boson mass and condensation
conditions. We further argue the robustness of this model over the wide range of parameters of mass, self coupling constant and the condensation rate. Then the
dynamics of BEC collapse and the preferred scale of the collapse are
studied. Finally, we describe possible observational tests of our model,
especially, the periodicity of the collapses and the gravitational wave
associated with them.
}

\begin{document}

\maketitle

\section{Introduction}

After WMAP, fundamental parameters in astrophysics, $H_{0}$, $q$, $\Omega
_{M},\;\Omega _{\Lambda },\;$etc. have became more accurate and mutually consistent.
Although the standard $\Lambda CDM$ model can explain most of the
observations, consistently with these parameters, it still remains as a
phenomenological model. In our previous paper \cite{F-M}, in order to
improve this model, we discussed a cosmological model in the framework of
relativistic Bose Einstein Condensate (BEC) and gave a partial solution of
"why now problem". There, the quantitative study has been limited within
spatially uniform development of condensates. In the present paper, we
develop this cosmological model to include spatially inhomogeneous modes to
describe the instability of BEC and to clarify the preferred scales of the
collapsed objects. Furthermore, we establish the robustness of this model,
especially showing a generality of the condensation strength.

In general, BEC proceeds in bose gas of mass $m$ and number density $n$,
when the thermal de Broglie wave length $\lambda _{dB}\equiv \sqrt{2\pi
\hbar ^{2}/(mkT)}$ exceeds the mean interparticle distance $n^{1/3}$, and
the wavepacket percolates in space,

\begin{equation}
kT<\frac{2\pi \hbar ^{2}n^{2/3}}{m}.  \label{eq2}
\end{equation}%
On the other hand, cosmic evolution has the same temperature dependence
since the matter dominant universe behaves, in an adiabatic process, as

\begin{equation}
\rho \propto T^{3/2}.  \label{eq5}
\end{equation}%
Hence if the boson temperature is equal to radiation temperature at $z=1000$%
, for example, we have the critical temperature at present $%
T_{critical}=0.0027\mbox{K}$, since $T_{m}\propto a^{-2}$ and therefore, 
$T_{\gamma}/T_{m}\propto a$ in an adiabatic evolution.$\ $Using the present energy
density of the universe $\rho =9.44\times 10^{-30} \mbox{g/cm}^{3}$, BEC takes place
provided that the boson mass satisfies

\begin{equation}
m<1.87 \mbox{eV}.\footnote{%
This constraint will be somewhat reduced later. See Eq.(32)}  \label{eq6}
\end{equation}%
\qquad \qquad \qquad\ \ 

Conventional BEC is described in terms of the mean field which obeys the
Gross-Pitaevskii (GP) equation

\begin{equation}
i\hbar \frac{\partial \psi }{\partial t}=-\frac{\hbar ^{2}}{2m}\Delta \psi
+V\psi +g\left\vert \psi \right\vert ^{2}\psi .  \label{eq3}
\end{equation}%
Here $\psi \left( {\vec{x},t}\right) $ is the condensate mean field, and $%
V\left( \vec{x}\right) $ is the potential. The coupling strength $g$ is
related with the s-wave scattering length $a$ as

\begin{equation}
g=4\pi \hbar ^{2}a/m,  \label{eq4}
\end{equation}

\noindent and therefore implies attractive interaction for $a<0$.
Originally, only positive value for $a$ (and therefore $g$) has been
considered for BEC, since negative value for $a$ necessarily yields
imaginary terms in the ground state energy and chemical potential\cite{L-P}.
However, the appearance of the imaginary part in the energy simply implies
that the ground state is unstable but BEC itself takes place, as a transient
state, even in negative $g$ \cite{Sackett}. This instability of BEC turns
out to be crucial in the context of cosmology.

The above GP equation is apparently non-relativistic. In the context of
cosmology, we need a relativistic GP equation. The relativistic GP equation
has a form of the Klein-Gordon equation with self-interaction
and the Lagrangian is given by 
\begin{equation}
L=\sqrt{-g}\left( {g^{\mu \nu }\partial _{\mu }\phi ^{\dag }\partial _{\nu
}\phi -m^{2}\phi ^{\dag }\phi -\frac{\lambda }{2}(\phi ^{\dag }\phi )^{2}}%
\right) .  \label{eq7}
\end{equation}%
We discuss the metric tensor given by 
\begin{equation}
\mathrm{d}s^{2}=(1+2\Phi )\mathrm{d}t^{2}-a^{2}(1-2\Phi )\mathrm{d}\mathbf{x}%
^{2}\,,  \label{metric2}
\end{equation}%
where $\Phi =\Phi \left( t,\vec{x}\right) $ represents the gravitational
potential. The instability of this field has already been studied in \cite%
{Khlopov} disregarding the cosmic expansion. We consider normal mass signature but the
self-coupling is negative. The instability analysis is applicable for either
signatures for $m^{2}$ and $\lambda $. We will discuss about the difference
of our formalism from the usual Higgs mechanism later.

In section 2, the essence of the BEC cosmology is briefly summarized. Scalar
dark matter plays an essential role in our scenario. We show in section 3
that BEC model is quite robust, especially the final accelerated-expansion
regime is always realized in wide range of parameters. In section 4,
extending the phenomenological scenario reviewed in section 2, we argue the
inhomogeneous modes and the instability of BEC based on the microscopic
Lagrangian Eq.(\ref{eq7}) with the metric Eq.(\ref{metric2}). Section 5 is
devoted to the miscellaneous observational problems to probe the remnants of
this BEC cosmological model. \ 

\section{Basics of BEC cosmology}

We briefly describe the basic scenario of the BEC cosmology developed in 
\cite{F-M,Nishiyama}. The backbones of this model are 1.
relativistic GP equation, 2. steady slow process of BEC, and 3. BEC
instability which leads to the Dark Energy collapse.

\subsection{ \ Relativistic Gross-Pitaevski equation}

From Eq.(\ref{eq7}), the relativistic version of the GP equation\footnote{%
We set $c=1,\;\hbar =1$ hereafter.}, in the Minkowsky space, becomes 
\begin{equation}
\frac{\partial ^{2}\phi }{\partial t^{2}}-\Delta \phi +m^{2}\phi +\lambda
(\phi ^{\ast }\phi )\phi =0,  \label{kg}
\end{equation}%
\noindent with the potential 
\begin{equation}
V\equiv m^{2}\phi ^{\ast }\phi +\frac{\lambda }{2}(\phi ^{\ast }\phi )^{2}.
\label{eq9}
\end{equation}%
Substituting the decomposition of the classical mean field $\phi =Ae^{iS}$
and defining the momentum $p_{\mu }=-\partial _{\mu }S=(\epsilon ,-\vec{p})$%
, where $\vec{p}=m\gamma \vec{v},~\gamma =\left( 1-\vec{v}^{2}\right)
^{-1/2} $, the relativistic GP equation reduces to the Euler equation for
fluid: 
\begin{equation}
\epsilon \frac{\partial \vec{v}}{\partial t}+\vec{\nabla}\left( \frac{\gamma
v^{2}}{2}+\frac{\lambda }{12m}A^{2}+\frac{\hbar ^{2}}{2Am}\partial _{\mu
}^{2}A\right) =0  \label{fluid}
\end{equation}%
The energy-momentum tensor
associated with Eq.(\ref{eq7}) becomes 
\begin{equation}
T_{\mu \nu }\equiv \frac{2}{\sqrt{-g}}\frac{\delta L}{\delta g^{\mu \nu }}%
=2\partial _{\mu }\phi ^{\ast }\partial _{\nu }\phi -g_{\mu \nu }(\partial
\phi ^{\ast }\partial \phi -m^{2}\phi ^{\ast }\phi -\frac{1}{2}(\phi ^{\ast
}\phi )^{2}).  \label{eq10}
\end{equation}%
For the isotropic relativistic fluid, it reduces to 
\begin{equation}
T^{\mu \nu }=\mathbf{diag}(\rho ,p,p,p),  \label{eq11}
\end{equation}%
\noindent in the local rest frame. Here, the condensate part of $\rho $ and $%
p$ are given by 
\begin{equation}
\rho =T^{00}=\dot{\phi}^{\ast }\dot{\phi}+m^{2}\phi ^{\ast }\phi +\frac{%
\lambda }{2}(\phi ^{\ast }\phi )^{2}=\dot{\phi}^{\ast }\dot{\phi}+V,
\label{eq12}
\end{equation}%
\noindent and 
\begin{equation}
p=T^{11}=T^{22}=T^{33}=\dot{\phi}^{\ast }\dot{\phi}-m^{2}\phi ^{\ast }\phi -%
\frac{\lambda }{2}(\phi ^{\ast }\phi )^{2}=\dot{\phi}^{\ast }\dot{\phi}-V.
\label{eq13}
\end{equation}

\subsection{Steady slow process of BEC}

We consider the cosmic evolution of various energy densities on average in
our model, leaving aside the dynamics of inhomogeneous modes, derived from
the microscopic Lagrangian (\ref{eq7}) to section 4. Further, here we
discuss the BEC cosmology phenomenologically, leaving aside the generality
of the parameters to section 3. The whole evolution is given by the
following set of equations\cite{F-M}. 
\begin{eqnarray}
H^{2} &=&\left( {\frac{\dot{a}}{a}}\right) ^{2}=\frac{8\pi G}{3c^{2}}\left( {%
\rho _{g}+\rho _{\phi }+\rho _{l}}\right) ,  \nonumber \\
\dot{\rho}_{g} &=&-3H\rho _{g}-\Gamma \rho _{g},  \nonumber \\
\dot{\rho}_{\phi } &=&-6H\left( {\rho _{\phi }-V}\right) +\Gamma \rho _{g}-{%
\Gamma }^{\prime }\rho _{\phi },  \nonumber \\
\dot{\rho}_{l} &=&-3H\rho _{l}+{\Gamma }^{\prime }\rho _{\phi }.
\label{eq16}
\end{eqnarray}%
Here $\Gamma $ is the decay rate of the boson gas (i. e. uniform DM) into
BEC, and $\Gamma ^{\prime }$ is the decay rate of BEC into collapsed BEC (i.
e. localized DM)). The former $\Gamma $ is a constant, but the latter $%
\Gamma ^{\prime }$ appears only when the BEC satisfies the instability
condition. These rates are transport coefficients which characterize the BEC
phase transition although they should be, in principle, calculated from the
Lagrangian Eq.(\ref{eq7}) and the environmental informations. Here we fix
these values phenomenologically in the present stage of our model\footnote{%
This is somewhat generalized later in section 3.2.}. As is easily seen from
Eq.(\ref{eq16}), DE$(\rho _{\phi })$ and DM($\rho _{g}+\rho _{l})$ are
intimately related with each other in our model; original uniform DM($\rho
_{g}$) condensates into DE$(\rho _{\phi }),$ and it collapses into localized
DM($\rho _{l}$), which eventually becomes a dominant component in the total
DM ($\rho _{g}+\rho _{l})$.

There are two relevant regimes of solutions to Eq.(\ref{eq16}). One is (a)
the over-hill regime, and the other is (b) the inflationary regime. The
former appears when the condensation speed is high and the condensation
overshoot the potential barrier, and the latter when it is low. A general
evolution is a mixture of them; several regimes of (a) finally followed by
the regime (b). Let us now examine these regimes.

\begin{figure}[tbp]
\resizebox{130mm}{!}{\includegraphics{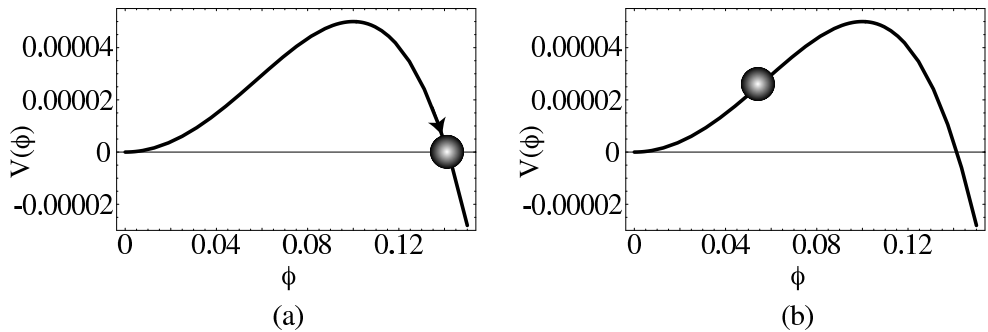}}
\caption{Schematic diagram of the evolution of BEC. (a) over-hill regime and
(b) inflationary regime are depicted. The inflation appears as a result of
the balance between the condensation and the potential force, $V^{\prime
}=\Gamma \protect\rho _{g}/\dot{\protect\phi}$.}
\label{fig:fig1}
\end{figure}

(a) \textit{Over-hill regime}: This regime generally appears when the
condensation strength is faster than the potential force and the
condensation overshoot the potential barrier, especially in the earlier
stage of the cosmic evolution. Actually, this regime is a fixed point of the
first three equations in Eq.(\ref{eq16}):

\begin{equation}
\phi \rightarrow \infty ,\rho _{\phi }\rightarrow 0,\rho _{g}\rightarrow
0,H\rightarrow 0,a\rightarrow a_{\ast }  \label{eq17}
\end{equation}%
The field goes over the hill of the potential, as in Fig.\ref{fig:fig1}(a).

The condensation speed $\Gamma $ is fast at the first stage, and the bose
gas density is simply reduced $\rho _{g}\propto e^{-\Gamma t}$ and the
cosmic friction term becomes negligible. Then, Eq.(\ref{eq16}c) yield $\ddot{%
\phi}\approx {V}^{\prime }$, and $\phi $ reaches singularity within a finite
time. Since $\dot{\phi}$ increases rapidly in the last stage of the fall,
the BEC reduction rate $-6H\left( {\rho _{\phi }-V}\right) \propto H\dot{\phi%
}^{2}$ dominates the BEC increase rate $\Gamma \rho _{g}$. Thus, we have
eventually $\rho _{\phi }\rightarrow 0$, and $H\rightarrow 0$. However,
actually, this virtual singularity is avoided by invoking the last equation
in Eq.(\ref{eq16}).

(b) \textit{Inflationary regime}: This regime appears when the condensation
strength is weaker than the potential force, especially in the later stage
of the cosmic evolution. This regime turns out to be a stable fixed point of
Eq.(\ref{eq16}): 
\begin{equation}
\phi \rightarrow \phi _{\ast },H\rightarrow H_{\ast },\rho \rightarrow 0,%
\dot{\phi}\rightarrow 0.  \label{eq25}
\end{equation}%
The BEC condensed field stops and stays at an intermediate position of the
potential hill for ever, as in Fig.\ref{fig:fig1}(b). This mechanism is a
novel type of inflation, which is supported by the balance of the
condensation speed ($\Gamma \rho _{g})$, and the potential force ($\dot{\phi}%
{V}^{\prime })$: 
\begin{equation}
\dot{V}=\Gamma \rho _{g}.  \label{eq26}
\end{equation}%
Though both sides of Eq.(\ref{eq26}) exponentially decay to zero, the
balance itself is automatically maintained\footnote{%
This exponentially decreasing amplitude of the balance may lead to the
instability of the inflationary regime and the autonomous termination of
this regime, given some small external perturbations.}.

In the actual universe, the above two kinds of regimes are realized
successively. First, the over-hill regime repeats multiple times in general
until the bose gas density is consumed and the condensation speed decreases.
Eventually the condensation force balances with the potential force, and the
final inflationary regime follows.

\begin{figure}[tbp]
\resizebox{130mm}{!}{\includegraphics{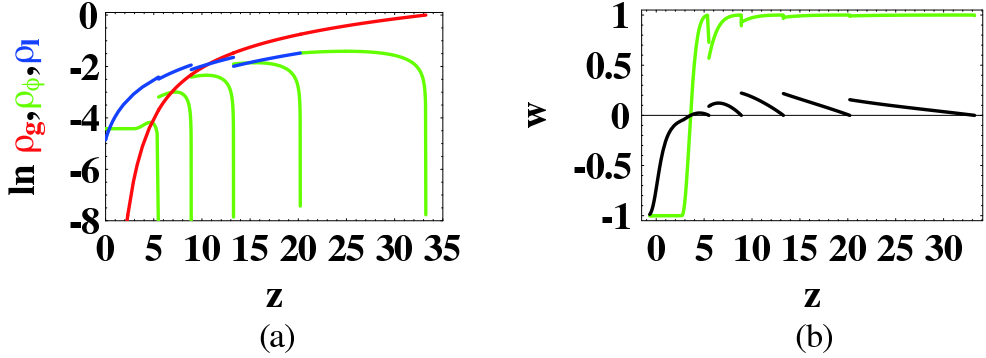}}
\caption{(a) Time evolution of various cosmic densities as a function of red
shift $z$. This is a numerical solution of Eq.(\protect\ref{eq16}). The red
line is $\protect\rho _{g}$, the green line is $\protect\rho _{\protect\phi %
},$\ and the blue line is $\protect\rho _{l}$. \ Here we have set $\tilde{m}%
^{2}\simeq 0.01,~\protect\lambda =-0.1,~\tilde{\Gamma}=0.4$. The variables
with tilde are dimensionless ones defined in the text. Several BEC collapses
take place, which finally followed by a phase with constant energy density
(accelerating expansion). (b) Time evolutions of the $w$\ parameters. The
green line represents $w_{\protect\phi }\equiv p_{\protect\phi }/\protect%
\rho _{\protect\phi }$ as a function of $z$. It is apparent that BEC behaves
as ideal gas in the early stage and as cosmological constant in the later
stage. Black solid line represents $w$ of the whole system. \ }
\label{fig:fig2}
\end{figure}

In Fig.\ref{fig:fig2}(a), numerical results for the evolution of the cosmic
energy densities are plotted. Black, dark gray, and light gray curves
represent, respectively in this order, the cosmic energy densities of the
bose gas ($\rho _{g}$), BEC ($\rho _{\phi }$), and the localized objects ($%
\rho _{l}$). In this example, four over-hill regimes are finally followed by
the inflationary regime.

Further, in Fig. \ref{fig:fig2}(b), the evolution of the corresponding EOS
parameter $w\equiv p/\rho $ is shown. It is clear that the BEC ($\rho _{\phi
}$) acquires the genuine DE property (i.e. $w\approx -1)$, only recently $%
z<3.$ For $z>3,$ $\rho _{\phi }$ behaves as ideal gas (i.e. $w\approx 1).$
This is because the field $\phi $ is in the stage of condensation and
moving. Therefore it possesses kinetic energy and positive pressure. Thus, a
genuine DE with $w\approx -1$ only appears for $z>3$ despite we often call $%
\rho _{\phi }$ as DE and $\rho _{g}+\rho _{l}$ as DM in this paper. Thus,
our model cannot be distinguished from the standard model with a
cosmological constant, as far as we observe the cosmic evolution $a(t)$.

\bigskip\ 

Now we explain the detail of our numerical calculations above and how to
obtain physical scales from them. In Eq.(\ref{eq16}) we simultaneously use
parameters of quite different orders of magnitudes, as a result of bridging
microphysics to the macrophysics, such as the plank mass $m_{Pl}$, cosmic
expansion rate $H$, and the boson mass $m$ etc. First we normalize
dimensional observables by two mass scales; $m_{\ast }$ for space-time, and $%
m_{0}$ for energy, whose orders are fixed later: 
\begin{equation}
\tau =m_{\ast }t,~~\tilde{H}=\frac{H}{m_{\ast }},~~\tilde{\rho}=\frac{\rho }{%
m_{0}^{4}},\;\widetilde{m}=\frac{m}{m_{0}},\;\widetilde{\phi }=\frac{\phi }{%
m_{0}},\;\widetilde{\Gamma }=\frac{\Gamma }{m_{\ast }},\;\widetilde{\Gamma }%
^{\prime }=\frac{\Gamma ^{\prime }}{m_{\ast }},
\end{equation}%
where $m_{\ast }$ and $m_{0}$ are related through the gravitational constant 
$G$ or plank mass $m_{Pl}:$ 
\begin{equation}
\frac{8\pi G}{3c^{2}m_{\ast }^{2}}=\frac{1}{m_{0}^{4}}\;\mbox{or}%
\;m_{0}^{2}=m_{pl}m_{\ast }
\end{equation}%
Then Eq.(\ref{eq16}) is rewritten as 
\begin{eqnarray}
\tilde{H}^{2} &=&\tilde{\rho _{g}}+\tilde{\rho _{\phi }}+\tilde{\rho _{l}} 
\nonumber \\
\tilde{\rho _{g}}^{\prime } &=&-3\tilde{H}\tilde{\rho _{g}}-\tilde{\Gamma}%
\tilde{\rho _{g}}  \nonumber \\
\tilde{\rho _{\phi }}^{\prime } &=&-3\tilde{H}\left( \frac{m_{\ast }}{m_{0}}%
\right) ^{2}\tilde{\phi}^{\prime \ast }\tilde{\phi}^{\prime }+\tilde{\Gamma}%
\tilde{\rho _{g}}-\tilde{\Gamma}^{\prime }\tilde{\rho _{\phi }}  \nonumber \\
\tilde{\rho _{l}}^{\prime } &=&-3\tilde{H}\tilde{\rho _{l}}+\tilde{\Gamma}%
^{\prime }\tilde{\rho _{\phi }},  \label{dl}
\end{eqnarray}%
where the prime means the derivative w.r.t. $\tau $. The ratio of the mass
scales $m_{\ast }/m_{0}$ appears only in the first term on the RHS of the
third equation of Eq.(\ref{dl}), and this term does not contribute since
this factor is extremely small, $m_{\ast }/m_{0}=m_{0}/m_{pl}\ll 1$.

The scales $m_{0}$ and therefore $m_{\ast }$ can be fixed, from our
numerical calculation, as follows. The energy density of BEC is normalized
as 
\begin{equation}
\tilde{\rho}_{\phi }=\left( \frac{m_{\ast }}{m_{0}}\right) ^{2}\tilde{\phi}%
^{\prime \ast }\tilde{\phi}^{\prime }+\tilde{m}^{2}\tilde{\phi}^{\ast }%
\tilde{\phi}+\frac{\lambda }{2}(\tilde{\phi}^{\ast }\tilde{\phi})^{2}.
\end{equation}%
We have started our calculation, in the above example of Fig.(\ref{fig:fig2}%
), with $\tilde{m}=0.1$ and obtained the numerical value for $\tilde{\rho}%
_{\phi }$ at present, $\tilde{\rho}_{\phi 0}=0.000015$. We identify $\rho
_{\phi 0}=0.73\rho _{cr0}$ with $\rho _{cr0}=9.44\times 10^{-30}$ g/cm$%
^{3} $, from which it follows that 
\begin{equation}
m_{0}\simeq 0.030\mbox{eV},~~m=\tilde{m}m_{0}=0.0030\mbox{eV},~~\mbox{and }%
m_{\ast }=2.09\times 10^{-31}\mbox{eV.}
\end{equation}

As we have mentioned after Eq.(\ref{eq16}), DE$(\rho _{\phi })$ and DM($\rho
_{g}+\rho _{l})$ are intimately related with each other in our model; DM($%
\rho _{g}$) condensates into DE$(\rho _{\phi }),$ and it collapses into
localized DM($\rho _{l}$). Therefore, it is natural to expect that the
amounts of DM and DE are almost the same. This is a kind of self-organized
criticality\cite{Jensen:1998} (SOC). Finally dominant DM($\rho _{l}$) is
composed from the quantity which had been DE($\rho _{\phi })$ before.

\subsection{BEC collapse\protect\bigskip}

There is no degenerate pressure for bosons unlike fermions. Therefore, only
the quantum pressure, expressed in the last term on the left-hand side of
Eq.(\ref{fluid}) , induced from the Heisenberg uncertainty principle can
prevent the BEC to collapse. However, there is a maximum mass for this
mechanism to work. The Compton wavelength of the object $\lambda
_{compton}=2\pi \hbar /\left( {mc}\right) $ must be larger than the
Schwarzschild black hole radius or its inner-most stable radius $3\left( {%
2GM/c^{2}}\right) $. This condition defines a characteristic mass scale 
\begin{equation}
M_{critical}\approx m_{pl}^{2}/m\equiv M_{KAUP}
\end{equation}%
\noindent only below which a stable configuration is possible for BEC. This
structure is well known as a boson star\cite{Kaup:1968}. For example, the
boson mass $m=10^{-3}\mbox{ eV}$ gives $M_{critical}=10^{-7}M_{\odot }$,
which is almost the mass of Mercury. If the object mass exceeds this
critical value, black holes are inevitably produced. There is no limit for
the black hole mass. Thus, DE black holes of any size are naturally produced
in our model. However actually, shock waves would also be naturally produced
in the collapse process of BEC. They convert the potential energy to heat
and yields huge pressure, which may prevent the collapse to black holes. We
will study the initial linear instability of BEC in the following sections.
In any case, it will be true that many compact clumps (boson stars, black
holes, hot clusters) are rapidly formed after the collapse of BEC. This
fascinating scenario, early-formation of black holes and clumps, has been
extensively discussed in \cite{KhlopovBH} in other context.

\section{Robustness of the BEC Cosmology}

After a brief introduction of BEC cosmology in the above, we would like to
argue first the robustness and generality of the BEC model. In particular,
we would like to clarify the condition under which the BEC phase is possible
in the universe. This argument is deeply related with the basic parameters
of the model: mass $m$ of the boson and the condensation rate $\Gamma $.

\subsection{Mass constraints and the BEC condition \ \qquad\ \ }

Let us now consider how BEC is possible in the expanding Universe. In
general, the charge density $n$ of the boson gas is expressed as the sum of
two contributions; the particle with positive signature and the
anti-particle with negative signature: 
\begin{equation}
n=\int \frac{dp^{3}}{(2\pi )^{3}}[\frac{1}{e^{\beta \left( \omega -\mu
\right) }-1}-\frac{1}{e^{\beta \left( \omega +\mu \right) }-1}],
\end{equation}%
which is a function of temperature $1/\beta $\ and the chemical potential $%
\mu $, and energy $\omega =\sqrt{p^{2}+m^{2}}$, in units of $\hbar
=k_{B}=c=1 $. BEC takes place when $\mu =m$ and the critical temperature $%
T_{c}$ or the critical density $n_{c}$\ is determined by setting so in the
above.

In the non-relativistic regime, i.e. $p^{2}\ll m^{2}$, the above form
yields, 
\begin{equation}
n_{c}=\zeta \left( 3/2\right) \left( \frac{mT}{2\pi }\right) ^{3/2}
\end{equation}%
or $T_{c}=(2\pi /m)\left( n/\zeta \left( 3/2\right) \right) ^{2/3}.$ Below $%
T_{c}$ or above $n_{c}$, the wave functions of individual particles begin to
overlap with each other, i.e. the thermal de Broglie length exceeds the mean
separation of particles, 
\begin{equation}
\lambda _{dB}\equiv \left( \frac{{2\pi }}{{mkT}}\right) ^{1/2}>r\equiv
n^{-1/3}
\end{equation}%
In this regime, the cosmic energy density of the non-relativistic matter has
the same dependence on the temperature: 
\begin{equation}
n=n_{0}\left( \frac{T}{T_{0}}\right) ^{3/2},
\end{equation}%
if we assume the entropy is conserved during the expansion and therefore $%
\rho \propto a^{-3}\propto T^{3/2}.$ Thus, provided that the cosmic
temperature had been once under the critical temperature at some moment in
the non-relativistic regime, the universe would be always under the critical
temperature and BEC can initiate all the time.

\begin{figure}[tbp]
\resizebox{100mm}{!}{\includegraphics{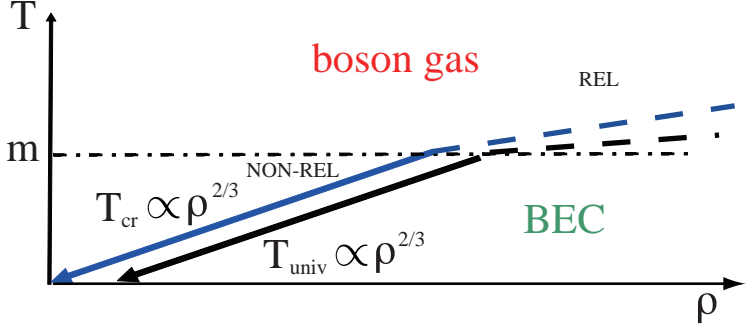}}
\caption{Schematic diagram of the critical temperature $T_{cr}$ and the
cosmic temperature $T_{universe}$. In the non-relativistic regime ($m<T$), $%
T_{cr}\propto \protect\rho ^{2/3},$ $T_{univ}\propto \protect\rho ^{2/3}$.
In the relativistic regime ($m>T$), $T_{cr}\propto \protect\rho ^{1/2},$ $%
T_{univ}\propto \protect\rho ^{1/4}$. The temperature evolution line for the
universe is the same as the adiabatic expansion. }
\label{fig:fig3}
\end{figure}

The above fact sets the upper limit on the boson mass, for BEC to take place
at the present universe. Let us suppose the boson gas was in thermal
equilibrium with radiation in the far past. Suppose the boson had
transformed from relativistic to non-relativistic at time $t_{\ast }$ with
the scale factor $a_{\ast }=a\left( t_{\ast }\right) $. The boson
temperature $T_{B}\left( t\right) $ at this time $t_{\ast }$ will be $%
T_{B}\left( t_{\ast }\right) =m=T_{\gamma }\left( t_{\ast }\right) $. After
that time, the boson temperature reduces as $\propto a^{-2}$, and the photon
temperature $T_{\gamma }\left( t\right) $\ as $\propto a^{-1}$. Therefore,
for \ $t>t_{\ast }$, 
\begin{equation}
T_{B}\left( t\right) =\left( \frac{a_{\ast }}{a\left( t\right) }\right)
^{2}m,~~T_{\gamma }\left( t\right) =\left( \frac{a_{0}}{a\left( t\right) }%
\right) T_{\gamma 0}.
\end{equation}%
Putting the present value of radiation temperature $T_{\gamma 0}=2.73K$ into
the above equation, we can estimate the present temperature of the boson gas 
$T_{B}\left( t_{0}\right) =T_{B0}$\ as 
\begin{equation}
T_{B0}=\left( \frac{T_{\gamma 0}}{m}\right) T_{\gamma 0}.
\end{equation}%
The present value of the critical temperature can be estimated from the
present energy density $\rho _{0}=9.44\times 10^{-30}g/cm^{3}$. The
ratio of them is 
\begin{equation}
\frac{T_{B0}}{T_{c0}}=\frac{\zeta \left( 3/2\right) ^{2/3}T_{\gamma
0}^{2}m^{2/3}}{2\pi \rho _{0}^{2/3}}.
\end{equation}%
The requirement that this ratio is smaller than $1$ yields the upper limit
of the boson mass: 
\begin{equation}
m<\frac{\left( 2\pi \right) ^{3/2}\rho _{0}}{T_{\gamma 0}^{3}\zeta \left(
3/2\right) }\approx 19 \mbox{eV}.  \label{mass ul}
\end{equation}%
It must be noted that this upper limit does not apply to the boson which has
not been in thermal equilibrium with radiation in the past, or the boson
composed from the pair of fermions.

On the other hand in the ultra-relativistic regime, i.e. $p^{2}\gg m^{2},$
the critical density becomes 
\begin{equation}
n_{c}=\frac{mT^{2}}{3},
\end{equation}%
and the cosmic energy density of the ultra-relativistic matter behaves as 
\begin{equation}
n=n_{0}\left( \frac{T}{T_{0}}\right) ^{4}.
\end{equation}%
Therefore, contrary to the nonrelativistic regime, even if the cosmic
temperature had been once under the critical temperature at some moment in
the ultra-relativistic regime, the boson temperature in the universe would
eventually goes over the critical temperature and BEC would melt into
thermal boson gas at that time. This trend is depicted in Fig.\ref{fig:fig3}.

\subsection{Time dependent $\Gamma $ and the robustness of the BEC model}

In the previous calculations \cite{F-M, Nishiyama}, we assumed the
condensation speed $\Gamma $\ is a constant parameter. However, this
quantity $\Gamma $ is not a simple term appearing in a Lagrangian, but a
transport coefficient, which includes all the information of the many-body
environment, e.g. temperature, density, fluctuations, etc. It should be
calculated from the quantum field theory of finite temperature and density
in the expanding universe, though such theory does not exist at present.
Therefore we take the second best method, i.e. we try all possible
time-dependent $\Gamma $. Although this does not specify $\Gamma $, we may
establish some robustness of the BEC cosmological model.

General transport coefficients would depend on the temperature and the
density of the environment. Time dependence of such global parameters can be
represented by a scale factor in the uniformly expanding universe.
Therefore, possible time dependences of the parameter $\Gamma $\ will be
exhausted by the inclusion of the scale factor, which normally depends in
the form of power. Thus we assume 
\begin{equation}
\Gamma =\bar{\Gamma}a\left( t\right) ^{\alpha }  \label{timedependentGamma}
\end{equation}%
where $\bar{\Gamma}$\ and $\alpha $\ are constants. Even in this case, the
basic mechanism of the BEC cosmology does not change. Actually, the
insertion of the expression Eq.(\ref{timedependentGamma}) in Eq.(\ref{eq16})
is equivalent to recast the behavior of boson gas density, which is the
source of condensation, as 
\begin{equation}
\rho _{g}\propto a\left( t\right) ^{-3}\rightarrow \rho _{g}\propto a\left(
t\right) ^{\alpha -3}  \label{recastrho}
\end{equation}%
while $\Gamma \rightarrow \bar{\Gamma}$\ is still a constant. Intuition
tells us that such change of the source gas density does not alter the
scenario in essence.

In order to check this intuition, we have performed several demonstrations
in numerical methods. Results are in Fig. 4.


\begin{figure}[tbp]
\resizebox{130mm}{!}{\includegraphics{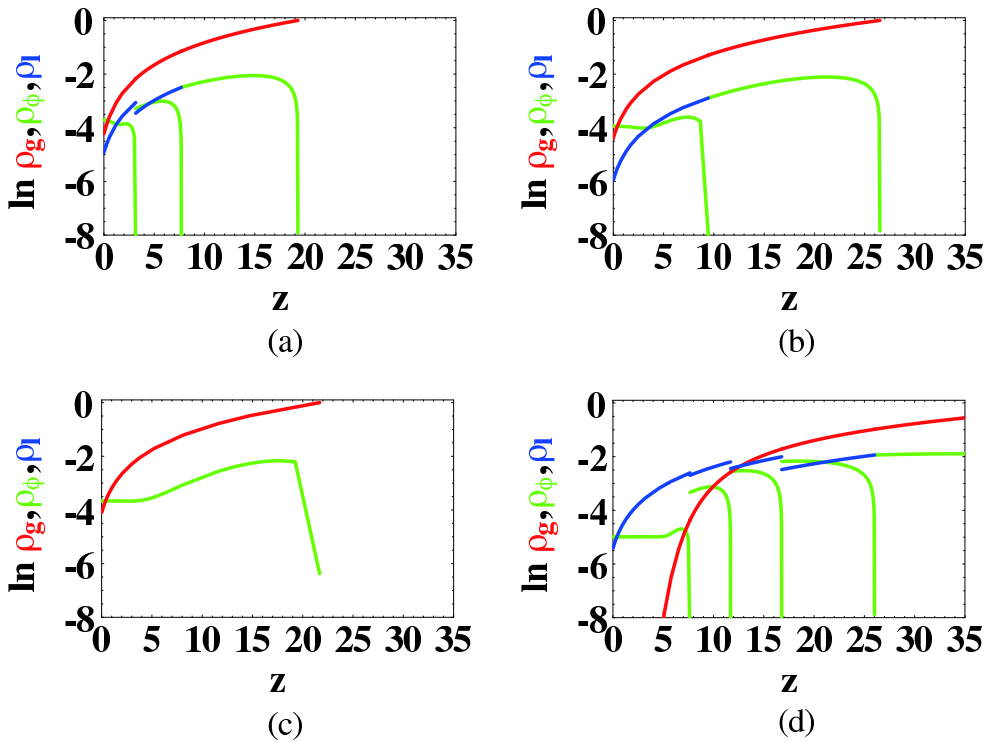}}
\caption{Time evolution of various cosmic densities as a function of red
shift $z$. Same as Fig 2(a), but with time dependent condensation rate $%
\tilde{\Gamma} = 0.1 a(t)^{\protect\alpha }$ where $\protect\alpha %
=-1,-2,-3$, and $+1$, respectively for (a), (b), (c), (d). Other parameters
are set as $\tilde{m}^{2}\simeq 0.01,~\protect\lambda =-0.1$. }
\label{fig:fig4}
\end{figure}


In general, the reaction rate reduces when the temperature decreases.
Therefore it will be natural to choose negative values for the parameter $%
\alpha $. In all calculations with $\alpha =-1,-2,-3$, the qualitative
behavior of the model, i.e. multiple BEC collapses followed by an inflation,
does not change. Quantitative changes are the total number of BEC collapse
and the identification of the present time. These results could be easily
foreseen from the fact that the change Eq.(\ref{recastrho}) is equivalent to
Eq.(\ref{timedependentGamma}). We have also performed positive values for $%
\alpha $ which may be less relevant in the actual universe.\ In this case,
only $\alpha =1$\ yields the qualitatively similar behavior, but the cases
for $\alpha >1$ are not clear within our calculations probably due to
numerical error.

\subsection{Numerical $\Gamma $ and $m$ -robustness of BEC}

Here we show the robustness of BEC on the relatively wide range of numerical
values of condensation rate $\Gamma $ and boson mass $m,$\ etc.

\begin{figure}[tbp]
\resizebox{130mm}{!}{\includegraphics{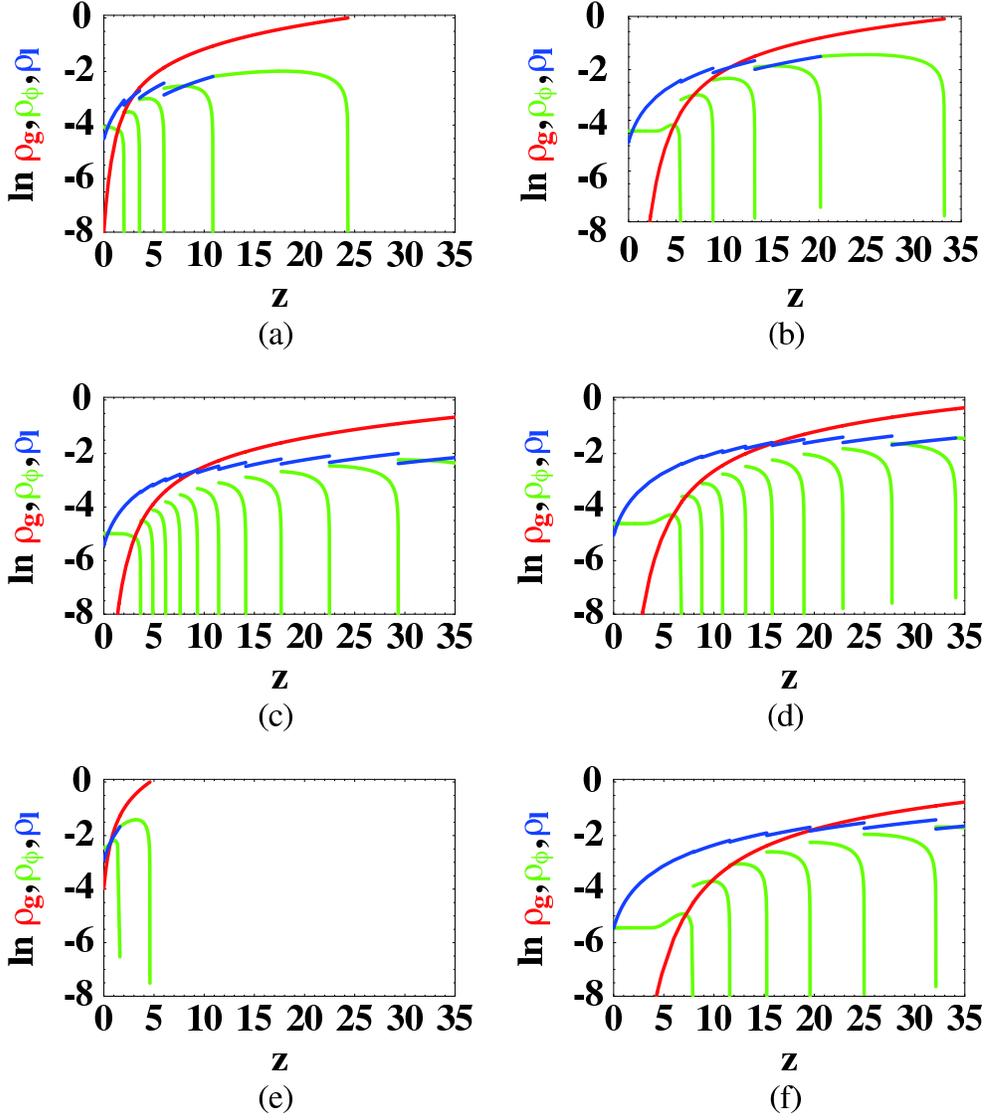}}
\caption{The robustness of BEC on the wide range of numerical values of $%
\widetilde{m},\;\widetilde{\Gamma }$, and $\protect\lambda $. (a) $%
\widetilde{m}^{2}=0.01,\widetilde{\Gamma }=0.1,\protect\lambda =-0.1$. (b) $%
\widetilde{m}^{2}=0.01,\widetilde{\Gamma }=0.4,\protect\lambda =-0.1$. (c) $%
\widetilde{m}^{2}=0.01,\widetilde{\Gamma }=0.1,\protect\lambda =-1$. (d) $%
\widetilde{m}^{2}=0.01,\widetilde{\Gamma }=0.4,\protect\lambda =-1$. (e) $%
\widetilde{m}^{2}=0.04,\widetilde{\Gamma }=0.4,\protect\lambda =-0.1$. (f) $%
\widetilde{m}^{2}=0.0025,\widetilde{\Gamma }=0.4,\protect\lambda =-0.1$. }
\label{fig:fig5}
\end{figure}

The boson mass turns out to be $0.0024\mbox{eV}$, $0.0030\mbox{eV}$,
 $0.0041\mbox{eV}$, $0.0033\mbox{eV}$, $0.0019\mbox{eV}$, $0.0024\mbox{eV}$,
respectively for the parameters (a)-(f) in Fig.\ref%
{fig:fig5}. The condensation rate $\Gamma $\ turns out to be $1.411\times
10^{-32} \mbox{eV}$, $8.37\times 10^{-32} \mbox{eV}$, 
$4.050\times 10^{-32}\mbox{eV}$,  $1.0521\times 10^{-31}\mbox{eV}$, 
$8.593\times 10^{-33}\mbox{eV}$, $2.272\times 10^{-31}\mbox{eV}$
in the same order. The qualitative feature of our model did not change under
the above variations of parameters. Further variation will be possible in
principle, but our numerical code at present does not yield reliable
results. This applies especially the late stage of the accelerating
expansion, due to the exponentially reducing energy density.

From the above results, we notice that the physical quantities $m$ and $%
\Gamma $ are not simply related with parameters $\tilde{m}$\ and $\tilde{%
\Gamma}.$ Actually, the boson mass $m$\ only changes about factor 2 while
the parameter mass $\tilde{m}$\ changes about factor 16. On the other hand,
the condensation rate $\Gamma $\ changes about factor 20 while the parameter 
$\tilde{\Gamma}$ only changes about factor 4.

Robustness of the boson mass value $0.003 \mbox{eV}\pm 0.001\mbox{eV}$
may be somewhat
interesting. However, we have not yet resolved the origin of this
robustness; it may be due to the intrinsic nature of our model, or it may
simply represent that our numerical range is too limited. Therefore, we will
not specify the boson mass in the following arguments, but leave some
possible rage of mass values open for further discussions.

\subsection{BEC instability and the collapse\protect\bigskip}

BEC collapse is a very complicated process. During the collapse of BEC,
gravitational potential energy, $GM^{2}/R,$ is released, where $M$ and $R$
are typical mass and scale of the collapsing region. If this collapse were
free from shock waves, the collapse were spherical symmetric and the
collapsing BEC exceeded the critical mass, then most of the collapsing BEC
would turn into black holes. However, such ideal conditions would never be
fulfilled.

If the BEC collapse takes place smoothly, and the process is adiabatic, then
the temperature of BEC is given by 
\begin{equation}
T\approx R^{3\left( 1-\gamma \right) }\approx M^{1-\gamma }\rho ^{\gamma -1},
\label{eq36b}
\end{equation}%
which is the same, if $\gamma =5/3,$ as the behavior of the adiabatic
universe in Fig.(\ref{fig:fig3}) , but the opposite time direction. The
boson temperature is always below the critical temperature, $T\propto \rho
^{2/3},$ and BEC is maintained.

If the BEC collapse takes place violently, and the gravitational energy
always turns into uniform thermal energy, then the process is in virtual
equilibrium, and the temperature of boson is given by 
\begin{equation}
T\approx \frac{GMm}{R}\approx GM^{2/3}m\rho ^{1/3},  \label{eq36}
\end{equation}%
which becomes $T\approx GMm/R\rightarrow m$, in the limit that the system
size approaches to the Schwarzschild radius $R\approx 2GM$. This means that
the boson becomes relativistic. Even in this case, the boson temperature is
always below the critical temperature, and BEC is maintained. See Fig.\ref
{fig:fig3}.

However in the real universe, shock waves are inevitably produced and
fluctuations associated with the collapse would be enormous. As a result,
some small portion of collapsed BEC terns into black holes and the rest
portion would become normal boson gas. In any case, the universe becomes
very clumpy at small scales.

The collapsed BEC will gravitationally attract baryons to form a cluster, as
in the standard CDM model. By contrast, at large scales, the potential is
not affected and is the same as that in the standard $\Lambda $CDM model.
This point is further clarified below.

We would like to restrict our study in the linear instability analysis in
this paper. This is the subject in the next section.

\section{Instability of BEC and large scale structure}

After examining robustness of the BEC cosmological model in the above, we
will now argue how the instability of BEC, which is extending on the entire
universe, can manifest to form localized structures.

The BEC collapse inevitably takes place in the over-hill regime in our
model. These collapsed components form localized objects and become direct
seeds of the structures in the universe. Our special interest is the
preferred scale of these structures. Though the full dynamics of the BEC
collapse would require involved numerical calculations, linear instability
analysis is always tractable, \cite{Khlopov}, within semi-analytical
calculations, to which we devote in this paper.

The metric is chosen as 
\begin{equation}
\mathrm{d}s^{2}=(1+2\Phi )\mathrm{d}t^{2}-a^{2}(1-2\Phi )\mathrm{d}\mathbf{x}%
^{2}\,,  \label{eqn:metric2}
\end{equation}%
where $\Phi =\Phi \left( t,\vec{x}\right) $ represents the gravitational
potential and $a=a\left( t\right) $ the cosmic scale factor. The Lagrangian
for the BEC condensate mean field $\phi $ becomes, on this metric without a
source term, 
\begin{eqnarray}
L &=&a(t)^{3}(1-2\Phi )\left( (1+2\Phi )^{-1}|\dot{\phi}|^{2}-a(t)^{-2}(1-2%
\Phi )^{-1}(\nabla \phi )\cdot (\nabla \phi ^{\dagger })\right.  \nonumber \\
&&\left. -m^{2}|\phi |^{2}-\frac{\lambda }{2}|\phi |^{4}+L_{g}\right) 
\nonumber \\
&\simeq &a(t)^{3}(1-4\Phi )\dot{\phi}^{2}-a(t)(\nabla \phi
)^{2}-m^{2}a(t)^{3}(1-2\Phi )\phi ^{2}  \nonumber \\
&&-\frac{\lambda }{2}a(t)^{3}(1-2\Phi )\phi ^{4}+(1-2\Phi )a(t)^{3}L_{g}\,,
\end{eqnarray}%
\bigskip where the last line is the linearized form. The source term can be
negligible in our analysis, since the BEC collapse takes place very rapidly
compared to the slow steady condensation time scale. The equation of motion
for the condensate becomes, up to the first order in $\Phi ,$ 
\begin{equation}
\ddot{\phi}+3\frac{\dot{a}}{a}\dot{\phi}-\frac{1}{a^{2}}\nabla ^{2}\phi
+m^{2}(1+2\Phi )\phi +\lambda (1+2\Phi )|\phi |^{2}\phi =0\,.
\label{eqn:scalar}
\end{equation}%
The factor $1+2\Phi $ in the last term, which was not in \cite{Khlopov},
plays an essential role. The associated Poisson equation becomes 
\begin{equation}
\nabla ^{2}\Phi =4\pi Ga^{2}\left\{ \dot{\phi}^{\dagger }\dot{\phi}+\frac{%
\nabla }{a}\phi \cdot \frac{\nabla }{a}\phi ^{\dagger }+m^{2}|\phi |^{2}+%
\frac{\lambda }{2}|\phi |^{4}-\rho _{0}\right\} ,  \label{eqn:Poisson}
\end{equation}%
where $\rho _{0}$ is the uniform background energy density.

We now decompose the variables into the background component, with suffices $%
0$, and the linearly perturbed component, with suffices $1$, as 
\begin{eqnarray}
\phi &=&\phi _{0}+\phi _{1}\,, \\
\Phi &=&0+\Phi _{1}\,.
\end{eqnarray}%
The background is spatially uniform and only depends on time, while the
perturbation is time and space-dependent. The background solution satisfies,
from Eqs.(\ref{eqn:scalar}), (\ref{eqn:Poisson}), 
\begin{eqnarray}
\ddot{\phi}_{0}+3\frac{\dot{a}}{a}\dot{\phi}_{0}+\left( m^{2}+\lambda |\phi
_{0}|^{2}\right) \phi _{0} &=&0\,,  \label{eqn:scalar-0th} \\
\dot{\phi}_{0}^{\dagger }\dot{\phi}_{0}+\left( m^{2}+\frac{\lambda }{4}|\phi
_{0}|^{2}\right) |\phi _{0}|^{2}-\rho _{0} &=&0\,.  \label{eqn:Poisson-0th}
\end{eqnarray}%
In these Eqs.(\ref{eqn:scalar-0th}), (\ref{eqn:Poisson-0th}), the variable $%
\phi _{0}$ can be assumed to be real without generality since all the
coefficients are real. The equations of motion for the perturbations are 
\begin{eqnarray}
0 &=&\ddot{\phi}_{1}+3\frac{\dot{a}}{a}\dot{\phi}_{1}-\frac{1}{a^{2}}\nabla
^{2}\phi _{1}+m^{2}(\phi _{1}+2\Phi _{1}\phi _{0})  \nonumber \\
&+&\lambda \left( \phi _{0}^{2}\phi _{1}^{\dagger }+2\phi _{0}^{2}\phi
_{1}+2\Phi _{1}\phi _{0}^{3}\right) \,,  \nonumber \\
\frac{1}{4\pi Ga^{2}}\nabla ^{2}\Phi _{1} &=&\dot{\phi}_{0}(\dot{\phi}_{1}+%
\dot{\phi}_{1}^{\dagger })+(m^{2}+\lambda |\phi _{0}|^{2})\phi _{0}(\phi
_{1}+\phi _{1}^{\dagger }).  \label{eq51}
\end{eqnarray}%
The variable $\phi _{1}$ is complex and $\Phi _{1}$ is real. Therefore we
replace them by three real functions: 
\begin{eqnarray}
\phi _{1} &=&x+iy\,, \\
\Phi _{1} &=&z\,,
\end{eqnarray}%
where $x=x\left( t,\vec{r}\right) $, $y=y\left( t,\vec{r}\right) $, and $%
z=z\left( t,\vec{r}\right) .$\ \ These new functions $x,y,z$ are decomposed
into Fourier modes%
\begin{eqnarray}
x &=&\bar{x}\exp \left( \Omega t+i\vec{k}\cdot \vec{r}\right) \,, \\
y &=&\bar{y}\exp \left( \Omega t+i\vec{k}\cdot \vec{r}\right) \,, \\
z &=&\bar{z}\exp \left( \Omega t+i\vec{k}\cdot \vec{r}\right) \,,
\end{eqnarray}%
where $\bar{x},\bar{y},\bar{z},\Omega ,\vec{k}$ are constants. Putting these
into Eqs.(\ref{eq51}), we have 
\begin{eqnarray}
\left( \Omega ^{2}+3\frac{\dot{a}}{a}\Omega +\frac{\mathbf{k}^{2}}{a^{2}}%
+m^{2}+3\lambda \phi _{0}^{2}\right) \bar{x}+2(m^{2}+\lambda \phi
_{0}^{2})\phi _{0}\bar{z} &=&0\,, \\
\left( \Omega ^{2}+3\frac{\dot{a}}{a}\Omega +\frac{\mathbf{k}^{2}}{a^{2}}%
+m^{2}+\lambda \phi _{0}^{2}\right) \bar{y} &=&0\,, \\
\left( 2\dot{\phi}_{0}\Omega +2(m^{2}+\lambda \phi _{0}^{2})\phi _{0}\right) 
\bar{x}+\frac{\mathbf{k}^{2}}{4\pi Ga^{2}}\bar{z} &=&0\,.
\end{eqnarray}%
Existence of a non-trivial solutions $\bar{x},\bar{y},\bar{z}$ requires that
the above set of linear equations are dependent with each other. Thus we
have the condition 
\begin{eqnarray}
&&\left[ \left( \Omega ^{2}+3\frac{\dot{a}}{a}\Omega +\frac{\mathbf{k}^{2}}{%
a^{2}}+m^{2}+3\lambda \phi _{0}^{2}\right) \frac{\mathbf{k}^{2}}{4\pi G}%
\right.  \nonumber  \label{det} \\
&&\left. ~~~~~~-2(m^{2}+\lambda \phi _{0}^{2})\phi _{0}\left( 2\dot{\phi}%
_{0}\Omega +2(m^{2}+\lambda \phi _{0}^{2})\phi _{0}\right) \right]  \nonumber
\\
&\times &\left( \Omega ^{2}+3\frac{\dot{a}}{a}\Omega +\frac{\mathbf{k}^{2}}{%
a^{2}}+m^{2}+\lambda \phi _{0}^{2}\right) =0\,,  \label{instability}
\end{eqnarray}%
which determines the instability parameter $\Omega $ as a function of the
wave number $\vec{k}.$ If one of the solution $\Omega $ in this equation
becomes positive for some $\vec{k}$, such mode becomes unstable with the
time scale $\Omega ^{-1}$\footnote{%
It should be remarked that even in the simple case $a=a_{0}=1,~\phi _{0}=$%
const, $\Omega ^{2}$ has positive real roots and is unstable for the case of 
$m^{2}<0,~\lambda >0$ unlike the statement in \cite{Khlopov}.}. Since the
shortest time scale for the structure of scale $l\equiv a/k$\ to form is $%
l/c $ from causality,\ such structure formation would be actually possible
only if $\Omega ^{-1}>l$. Thus the structure of linear scale $l$ is possible
only if the condition $k/a>\Omega $ is satisfied. More precisely, the first
structure formation takes place at the shortest time scale. This condition,
by setting $\alpha $ as a positive constant smaller than unity,%
\begin{equation}
\alpha \frac{k}{a}=\Omega \mbox{ with }0<\alpha <1,\   \label{k=omega-cond}
\end{equation}%
uniquely determines the preferred linear scale $a/k_{\ast }$ of the
structure formed after the BEC collapse.

The expression for $\alpha k/a=\Omega $ is solved for $k$, choosing the
value associated with the most unstable mode among four solutions of $\Omega 
$ for Eq.(\ref{det}), as%
\begin{equation}
\frac{k_{\ast }}{a}=\left( \frac{-m_{eff}^{2}+\sqrt{m_{eff}^{4}+64\pi
G(1+\alpha ^{2})\left( m^{4}\phi _{0}^{2}-2\kappa m^{2}\phi _{0}^{4}+\kappa
^{2}\phi _{0}^{6}\right) }}{2(1+\alpha ^{2})}\right) ^{1/2}
\label{solk=omega}
\end{equation}%
where adiabatic approximation $H=0,$\ $\dot{\phi}_{0}=0$\ is utilized since the collapse time scale is much smaller than cosmic and condensation time scales. 
In Eq.(\ref{solk=omega}), we consider the regime
so that $m_{eff}^{2}\equiv m^{2}-3\kappa \phi _{0}^{2}>0$, i.e. $\phi _{0}$
smaller than the value at the inflection point of the potential $V\left(
\phi \right) $. This regime is first realized in the BEC condensation
process. The above expression Eq.(\ref{solk=omega}) can be further reduced
to 
\begin{equation}
\frac{k_{\ast }}{a}\approx \frac{8m^{2}\phi _{0}}{m_{eff}}\sqrt{\pi G\left(
1-2\kappa \left( \frac{\phi _{0}}{m}\right) ^{2}+\kappa ^{2}\left( \frac{%
\phi _{0}}{m}\right) ^{4}\right) },  \label{k=omega red}
\end{equation}%
since in general, the present mass scales, i.e. of order $eV$, are
negligibly small in comparison with the Plank mass $10^{28}eV$: $%
m_{eff}^{2}\approx \kappa \phi _{0}^{2}\approx m^{2}\ll m_{pl}^{2}\approx
G^{-1}$.

A rough estimate of Eq.(\ref{k=omega red}) and of the preferred scale is
possible. Setting $m_{eff}^{2}\approx \phi _{0}^{2}\approx m^{2},$ Eq.(\ref%
{k=omega red}) yields the linear scale $l_{\ast }\equiv a/k_{\ast }$ 
\begin{equation}
l_{\ast }\approx \left( \frac{m_{pl}}{m}\right) \frac{1}{m}\approx
10^{23} \mbox{cm}\approx 30\left( \frac{\mbox{eV}}{m}\right) ^{2} \mbox{kpc}.
  \label{1/k}
\end{equation}%
The resultant mass scale associated with the BEC collapse, which took place
at redshift $z$,\ $\ $would be 
\begin{equation}
M_{\ast }=\rho _{0}z^{3}\frac{4\pi }{3}l_{\ast }^{3}\approx 1.6\times
10^{11}\left( \frac{z}{20}\right) ^{3}\left( \frac{m}{\mbox{eV}}\right)
^{-6}M_{\odot },  \label{M*}
\end{equation}%
in which strong mass dependence is apparent. Actually there are several
different scenarios for the formation of localized structures, depending on
the mass $m$ of the boson. We suppose the first BEC collapses at about
redshift $z\approx 20.$

(a) If the boson mass is about $1 \mbox{eV}$, then, the typical mass of the
structure will be 
\begin{equation}
M_{\ast }\approx 1.6\times 10^{11}M_{\odot }  \label{M*eV}
\end{equation}%
which is of order of a galaxy. Some fraction of this mass turns into a black
hole and the remaining boson becomes a hot gas surrounding the black hole.
This is the typical structure expected from the BEC collapse in the present
cosmological model for this boson mass. The detail of the mass of such black
holes necessitates elaborate numerical calculations, on which we will report
in a separate report in the future.

(b) If the boson mass is about $10^{-3} \mbox{eV}$,
then the typical scale well
exceeds the size of the horizon, as easily observed from the power in Eq.(%
\ref{M*}). Thus, a structure is not formed in the early stage of the BEC
condensation while the condition $m_{eff}^{2}\equiv m^{2}-3\kappa \phi
_{0}^{2}>0$ holds. In this case, BEC condensation further proceeds and
crosses over the inflection point, beyond there $m_{eff}^{2}$ becomes
negative and we can no longer use the approximation $m_{eff}^{2}\approx \phi
_{0}^{2}\approx m^{2}$. Then, we have to go back to Eq.(\ref{solk=omega}),
which yields the solution%
\begin{equation}
\frac{k_{\ast }}{a}\approx \frac{\left\vert m_{eff}\right\vert }{\sqrt{%
1+\alpha ^{2}}}.  \label{k=omega red2}
\end{equation}%
It means that the strong instability initiates immediately after the mean
field crosses over the inflection point. Thus we define the time $\tau $ as
the elapsed time after crossing the inflection point. A structure of scale $%
a/k$ is formed at around $\tau =a/(\alpha k).$ Since the BEC is very
unstable and the time scale is short, we can expand $\phi \left( \tau
\right) =\phi _{\inf }+\dot{\phi}_{\inf }\tau +O\left( \tau ^{2}\right) $,
where $\phi _{\inf }$ is the value of the condensation at the inflection
point: i.e. $m^{2}=3\kappa \phi _{\inf }^{2}.$ Utilizing the relations, ($%
k_{\ast }/a)^{2}\approx \left\vert m^{2}-3\kappa \phi ^{2}\right\vert
/(1+\alpha ^{2})$ $=$ $6\kappa \phi _{\inf }\dot{\phi}_{\inf }\tau
/(1+\alpha ^{2}),$ we have $\tau ^{-1}=(6\kappa \dot{\phi}_{\inf }\phi
_{\inf }\alpha ^{2}/(1+\alpha ^{2}))^{1/3}$, and therefore $k/a=(\alpha \tau
)^{-1}=(6\kappa \dot{\phi}_{\inf }\phi _{\inf }/(\alpha +\alpha ^{3}))^{1/3}$%
. Putting approximate values $\dot{\rho}_{\phi }=-6H\left( {\rho _{\phi }-V}%
\right) +\Gamma \rho _{g}\approx \Gamma \rho _{g}$, and $\rho _{g}\approx
\rho _{\phi }\approx m^{2}\phi ^{2}$, we have 
\begin{equation}
\frac{k_{\ast }}{a}\approx (6\kappa \phi ^{2}\Gamma /(\alpha +\alpha
^{3}))^{1/3}\approx (m^{2}\Gamma )^{1/3}\approx
((10^{-3} \mbox{eV})^{2}10^{-32} \mbox{eV})^{1/3}\approx 10^{-13} \mbox{eV},
  \label{k-smallmass}
\end{equation}%
which corresponds to the size $l_{\ast }\approx 10^{3} \mbox{km}$ and
$M_{\ast}\approx 10^{-1} \mbox{g}.$ Therefore, BEC collapse cannot
form macroscopic
cosmological structures. These clumps would work as the ordinary DM and the
scenario of large structure formation reduces to the standard model.

Thus the above situations can be summarized as follows. We have two Jeans
wave numbers: one is $m^{2}/m_{pl},$ which appears for the condensation $%
\phi _{0}$\ smaller than the inflection point, and $(m^{2}\Gamma )^{1/3},$\
which appears for $\phi _{0}$ larger than the inflection point. For the case
(a) $m\approx 1 \mbox{eV}$,
the former instability takes place and the latter has no
chance to appears. For the case (b) $m\approx 10^{-3} \mbox{eV}$, the former
instability is not sufficient and the latter strong instability sets in to
make the BEC collapse.

(c) If the boson mass is far below $10^{-3} \mbox{eV}$,
then we can estimate the
preferred mass scale utilizing the above scaling relation $M\propto m^{-2}$.
For example, the boson of mass $10^{-22} \mbox{eV}$
would yield an object of the
galaxy size, and $10^{-24} \mbox{eV}$ a cluster size. Because the mass
is ultra-low,
the boson would always be in the condensed phase. Therefore the BEC has a
chance to form DM directly. This consideration naturally brings us to a
popular idea that the DM around a galaxy or a cluster is formed from scalar
field with ultra-low mass \cite{Mielke2006,Guzman2006,Silverman}. We will leave this possibility open in this paper, and proceed
to the next subject; radiation of gravitational wave.

\section{Observational remnants of the BEC cosmology}

We now turn our attention to possible observational remnants of the BEC
cosmological history, especially in the context of BEC collapses. Most
prominent effect would be the emission of gravitational wave, which may be
remaining as a fossil in our present universe.

\subsection{Gravitational wave associated with the BEC decay}

In our model, universe repeats violent decay of BEC to localized objects in
general. Associated with this process, gravitational wave is expected to be
produced. The energy emission rate of the gravitational wave from the moving
body can be calculated from the formula

\begin{equation}
\frac{dE}{dt}=\frac{G}{45c^5}\left(\frac{d^3D_{\alpha \beta }}{dt^3}\right)^2
\label{eq31}
\end{equation}
where

\begin{equation}
D_{\alpha \beta }=\int {\rho \left( {3x_{\alpha }x_{\beta }-r^{2}\delta
_{\alpha \beta }}\right) dV}  \label{eq32}
\end{equation}%
is the quadrupole of the whole mass distribution. Suppose the object of
linear size $R$ collapses with the typical speed $\ v$. Then the total
energy emitted would be the integration of the above formula during the
collapsing time scale,

\begin{equation}
E\approx \frac{dE}{dt}\Delta t\approx \left( {\frac{GM^{2}}{R^{2}}\frac{v^{6}%
}{c^{5}}}\right) \left( {\frac{R}{v}}\right) =\frac{GM^{2}}{R}\left( {\frac{v%
}{c}}\right) ^{5}.  \label{eq33}
\end{equation}%
This is roughly the gravitational potential energy of the extended object
multiplied by the 'efficiency' $\left( v/c\right) ^{5}$. We now estimate
possible remnant gravitational wave in the present background sky, for
several cases bellow.

(a) If we adopt the mass of the boson is about $1 \mbox{eV}$,
 then from Eq.(\ref%
{M*eV}), the preferred scale is $R=10^{23} \mbox{cm}$, \ $M=1.6\times
10^{11}M_{\odot }$. If we tentatively assume $v=c/10$, then

\begin{equation}
E\approx 10^{53} \mbox{erg}  \label{eq34}
\end{equation}%
which should be compared with the total rest energy of a star: $M_{\odot
}=10^{54} \mbox{erg}$. If we further assume that the first BEC collapse
took place
at redshift $z\approx 20$, then the present energy density of the
gravitational wave becomes

\begin{equation}
\rho _{gr,z=0}=\rho _{gr,z=20}\left( {20}\right) ^{-4}=\frac{E}{R^{3}}\left( 
{20}\right) ^{-4}=10^{-21}\frac{\mbox{erg}}{\mbox{cm}^{3}}  \label{eq35}
\end{equation}%
which should be compared with the total energy density at present, $\rho
_{cr}=10^{-29} \mbox{g}/\mbox{cm}^{3}=10^{-8} \mbox{erg}/\mbox{cm}^{3}$,
thus 
\begin{equation}
\Omega _{gr,z=0}=10^{-13}.  \label{eq35.5}
\end{equation}

Since the strain $h$ associated with the gravitational wave is related with

\begin{equation}
\Omega _{gw}=\frac{\omega ^{2}h^{2}}{12H_{0}^{2}},  \label{eq37}
\end{equation}
where $\omega $\ is the frequency of the wave, we have

\begin{eqnarray}
h &\approx &10^{-11}\mbox{ for }\omega \approx \left( {30 \mbox{kpc}}\right)
^{-1}=10^{-12} \mbox{Hz},  \label{eq38} \\
h &\approx &10^{-26}\mbox{ for }\omega \approx 10^{3} \mbox{Hz},
\end{eqnarray}%
where the frequency is estimated by naive extrapolation. This should be
compared with the present limit of the gravitational wave $h\approx 10^{-21}$
for $\omega \approx 10^{3} \mbox{Hz}$.

The gravitational background formed during the inflation is estimated\cite%
{peacock} as

\begin{equation}
\Omega _{gw}=\left( {H/m_{pl}}\right) ^{2}\Omega _{r}\approx \left( {10^{-5}}%
\right) ^{2}10^{-4}=10^{-14},  \label{eq39}
\end{equation}%
and the associated strain is

\begin{equation}
h\approx 10^{-27}\quad \mbox{for}\quad \omega \approx 10^{3} \mbox{Hz}.
  \label{eq40}
\end{equation}%
The gravitational background formed during the oscillation of the cosmic
string\cite{peacock} is

\begin{equation}
\Omega _{gw}=100\left( {G\mu /c^{2}}\right) \Omega _{r}\approx
100 \cdot 10^{-5} \cdot 10^{-4}=10^{-7}.  \label{eq41}
\end{equation}

(b) If we adopt the mass of the boson is about $10^{-3} \mbox{eV}$,
 then from Eq.(%
\ref{k-smallmass}), the preferred scale is $R=10^{8} \mbox{cm}$,
 $M=10^{-1} \mbox{g}$.
Then the present energy density of the gravitational wave becomes

\begin{equation}
\rho _{gr,z=0}=10^{-51}\frac{\mbox{erg}}{\mbox{cm}^{3}},\;
\Omega _{gr,z=0}=10^{-43},
\end{equation}%
mainly at frequency $\omega \approx \left( 10^{8} \mbox{cm}\right)^{-1}
\approx 10^{3} \mbox{Hz},$ which is totally small and would be never
 be detected.

The overall amount of energy density, as estimated in Eq.(\ref{eq35.5}),
will not affect the present standard cosmology. However, the strain, as
estimated in Eq.(\ref{eq38}), may have a chance to be detected as well as
the case of gravitational wave formed in the inflationary stage.

\subsection{log-z periodicity}

The BEC collapses do not take place randomly but they are periodical events
in the logarithm of cosmic time. As argued in section 2, the bose gas
density is simply reduced as $\rho _{g}\propto e^{-\Gamma t}$ in the
over-hill regime since the condensation speed $\Gamma $ is faster than the
cosmic dilution time scale. Therefore $\rho _{g}$ is simply transformed into 
$\rho _{\phi }.$ Just after each collapse, new BE-condensation always begins
from $\phi =0$ until it reaches some critical value $\rho _{\phi
}^{cr}=O\left( 1\right) V_{\max }\approx m^{4}/\left( {-\lambda }\right) $.
Therefore the condensation energy density behaves $\left[ {\rho _{g}\left( {%
t_{0}}\right) -\rho _{g}\left( t\right) }\right] $ modulo $\rho _{\phi
}^{cr} $ and 
\begin{equation}
\rho _{\phi }\left( t\right) \approx \left[ {\rho _{g}\left( {t_{0}}\right)
-\rho _{g}\left( t\right) }\right] _{\mbox{mod}\;\rho _{\phi }^{cr}}\approx %
\left[ {\rho _{g}\left( {t_{0}}\right) \left( {1-e^{-\Gamma t}}\right) }%
\right] _{\mbox{mod}\;\rho _{\phi }^{cr}},  \label{eq20}
\end{equation}%
\noindent where $t_{0}$ is the time when the first condensation begins.%

Accordingly, we expect that each BEC collapse takes place after the time
interval $\Delta t$ from the preceding collapse at time $t$, where $\Delta t$
is determined by the condition 
\begin{equation}
\rho _{\phi }^{cr}=\rho _{g}\left( {t_{0}}\right) \left( {e^{-\Gamma
t}-e^{-\Gamma \left( {t+\Delta t}\right) }}\right)  \label{eq21}
\end{equation}%
This implies that the occurrence of the BEC collapse is periodic in the
logarithm of time, $\log (t)$. If the cosmic expansion is power law in time,
i.e. $a\left( t\right) \propto t^{\mbox{const}}$, then $\log $($t$)-
periodicity directly implies log($a$) and log($z$) -periodicities. For
example, in the typical numerical calculation in Fig.(\ref{fig:fig2}), BEC
collapse takes place at $z=\{33.2,20.2,13.3,8.89,5.50\}$, which is almost
log-periodic.

This log($z$)-periodicity is a general consequence of our model, provided
that the over-hill regime and BEC repeats several times. Furthermore, this
periodic BEC collapse may leave its trace in the non-linear regime in forms
such as the discrete scale invariance or the hierarchical structure in the
universe, provided the scale is appropriately chosen.

As was discussed in the previous section, the preferred scale of BEC
collapse only depends on the basic parameters of the model; $l_{\ast
}\approx m_{pl}/m^{2}$ for $m>1 \mbox{eV}$ and
$\ \ l_{\ast }\approx (m^{2}\Gamma )^{1/3}$ for $m<10^{-3}\mbox{eV}$.
Then the sequence of the cluster produced
through BEC collapses has the series of mass proportional to $z^{3}$, where $%
z$\ is the redshift of the collapse. Thus, the cluster mass also has the
log-periodicity, and the largest cluster is formed at the first BEC
collapse. The most interesting case would be when the boson mass is about $%
1 \mbox{eV}$. Then the hierarchy of galaxies are formed, for example
 in Fig.(\ref{fig:fig2}), $M/M_{\odot }=\{1.6\times 10^{11}$,
 $3.6\times 10^{10}$, $%
1.0\times 10^{10}$, $3.1\times 10^{9},7.2\times 10^{8}\}$. Detail of the
arguments on the observational size is a future problem. If the boson mass
is about $10^{-3} \mbox{eV}$, then the hierarchy would yield no
interesting scales
with respect to the large scale structure formation.

\section{Summary}

We have developed the cosmological model based on the Bose-Einstein
Condensation (BEC) from various points of view. This BEC cosmology is
characterized by (1) the unification of DE and DM, (2) their mutual
conversion, (3) quantum mechanical condensation as a novel phase of DE, (4)
violent collapse of DE, (5) log-z periodicity of the DE collapses, (6) black
hole formation from DE, (7) formation of localized objects in high redshift
regime, (8) inevitable final phase of accelerated expansion, etc. These have
been briefly explained in section 2. We have examined this model in detail
especially with respect to its \textsl{robustness and instability} in this
paper.

We have first examined the \textsl{robustness} of the model in section 3. By
assuming thermal equilibrium of the boson field and the radiation in the
early universe, we set the upper limit of the mass for the boson, which
turned out to be about $\mbox{eV}$. Thus we have showed that BEC takes
place very
naturally in the universe. Moreover, if the boson has not been in
equilibrium with radiation in the past, then even any value of mass is
allowed. Next we demonstrated that the BEC takes place in the wide range of
microscopic parameters of boson mass $m$, self coupling $\lambda $, and the
condensation rate $\Gamma $. Furthermore, we have revealed that the time
dependence of $\Gamma $\ does not qualitatively affect the BEC model. Thus
we have been able to show the robustness and the naturalness of the cosmological BEC
model.

Then we have examined the \textsl{instability} of BEC in section 4. We have
calculated a preferred scale of the structure formed after the BEC collapse
associated with this instability. This scale turns out to be quite sensitive
to the mass of the boson. If the boson mass is about $1 \mbox{eV}$,
 the preferred
scale is $l_{\ast }\approx m_{pl}/m^{2}$ and it is about a galaxy size. If
the boson mass is about $10^{-3} \mbox{eV}$, the scale is $\ l_{\ast }\approx
(m^{2}\Gamma )^{1/3}$ and it is about a gram. If the boson mass is far much
smaller, there is a possibility that DM is formed as BEC, and the preferred
scale can be of galaxy or cluster size. We have also estimated the amount of
the remnant gravitational wave associated with the BEC collapse. It turns
out to be marginally observable in the near future if the parameters of our
model is most optimized, otherwise it is simply too small.

The present results as a whole, suggest that the boson mass is probably of
order $10^{-3} \mbox{eV}$ to $1\mbox{eV}$.
These mass scales are so tiny that no empirical
evidence for such boson has been found yet. However, it is potentially
interesting that these mass scales are the same order as the neutrino
masses. Therefore it would be natural to consider that the boson particle is
a composite of neutrino-neutrino(or neutrino antineutrino) pair
\cite{yabu,capolupo}
 though this requires further investigations including the
problem of how Fermi surface can be stable in such tiny mass. We would like
to further extend the cosmological BEC model in our next paper. \ 

\section*{Acknowledgements}

The work of T.F is supported by Grant-in-Aid for Scientific Research
from the Ministry of Education, Science and Culture of Japan (\#16540269). He is also 
grateful to C.S.Kim for his hospitality at Yonsei
University and to BK21 of Korean Government for support of stay at Yonsei
University.

\end{document}